# Interplay between structural and magnetic-electronic responses of $FeAl_2O_4$ to a megabar: site inversion and spin crossover

W. M. Xu,[1#] G. R. Hearne,[2] S. Layek,[1] D. Levy,[1] M. P. Pasternak,[1]
G. Kh. Rozenberg[1] and E. Greenberg[1,3]

[1]*School of Physics and Astronomy, Tel Aviv University, 69978 Tel Aviv, Israel*

[2]*Department of Physics, University of Johannesburg, P.O. Box 524 , Auckland Park, 2006, Johannesburg, South Africa*

[3]*Center for Advanced Radiation Sources, University of Chicago, Argonne, IL 60439, USA*

## Abstract

X-ray diffraction pressure studies at room temperature demonstrate that the spinel $FeAl_2O_4$ transforms to a tetragonal phase at ~18 GPa. This tetragonal phase has a highly irregular unit-cell volume versus pressure dependence up to ~45 GPa, after which a transformation to a *Cmcm* post-spinel phase is onset. This is attributable to pressure driven Fe ↔ Al site inversion at room temperature, corroborated by signatures in the $^{57}$Fe Mössbauer spectroscopy pressure data. At the tetragonal → post-spinel transition, onset in the range 45–50 GPa, there is a concurrent emergence of a non-magnetic spectral component in the Mössbauer data at variable cryogenic temperatures. This is interpreted as spin crossover at sixfold coordinated Fe locations emanated from site inversion. Spin crossover commences at the end of the pressure range of the tetragonal phase and progresses in the post-spinel structure. There is also a much steeper volume change $\Delta V/V$ ~ 10% in the range 45–50 GPa compared to the preceding pressure regime, from the combined effects of the structural transition and spin crossover electronic change. At the highest pressure attained, ~106 GPa, the Mössbauer data evidences a diamagnetic Fe low-spin abundance of ~50%. The rest of the high-spin Fe in eightfold coordinated sites continue to experience a relatively small internal magnetic field of ~33 T. This is indicative of a magnetic ground state associated with strong covalency, as well as substantive disorder from site inversion and the mixed spin-state configuration. Intriguingly magnetism survives in such a spin-diluted post-spinel lattice at high densities. The *R*(300 K) data decreases by only two orders of magnitude from ambient pressure to the vicinity of ~100 GPa. Despite a ~26% unit-cell volume densification from the lattice compressibility, structural transitions and spin crossover, $FeAl_2O_4$ is definitively non-metallic with an estimated gap of ~400 meV at ~100 GPa. At such high densification appreciable bandwidth broadening and gap-closure would be anticipated. Reasons for the resilient non-metallic behavior are briefly discussed.

[#] Author to whom all correspondence should be addressed : xuw@post.tau.ac.il

## I. INTRODUCTION

Hercynite $FeAl_2O_4$ is a 2-3 cubic spinel, which is isomorphic to the mineral $MgAl_2O_4$. In this structure the $Fe^{2+}$ and $Al^{3+}$ cations can occupy both tetrahedral *A* and octahedral *B* sites of a spinel with a general compound formulation $AB_2O_4$. An inversion parameter *x* specifying the fraction of *A* sites occupied by trivalent cations in hercynite leads to a formulation: $\left(\mathrm{Fe}^{2+}_{1-x}\mathrm{Al}^{3+}_{x}\right)^{\mathrm{IV}}\left(\mathrm{Fe}^{2+}_{x}\mathrm{Al}^{3+}_{2-x}\right)^{\mathrm{VI}}\mathrm{O}_4$ , where the roman numerals refer to the fourfold and sixfold coordination sites. The degree of inversion at ambient conditions depends on synthesis conditions, viz, high temperature annealing value and cooling rate [1, 2]. The cation distribution amongst *A* and *B* sites in turn determines the chemical and physical properties of the spinel.

From a magnetic perspective the normal spinel $x = 0$ case, $FeAl_2O_4$, is of considerable interest. The *B*-site cations form a pyrochlore sublattice. *A*-site cations in these ferrous spinels form a diamond sublattice. This in itself does not lead to geometric spin frustration, but nearest-neighbor spin coupling $J_1$ and next nearest neighbor coupling $J_2$ engenders strong frustration. Consequentially exotic spin liquid or orbital liquid/glass ground states may be stabilized, as ascertained from comprehensive magnetic susceptibility and neutron scattering experiments [3, 4]. Different degrees of inversion disrupt these exotic ground states and for relatively low levels of inversion, say $x \sim 0.2$, spin glass behavior tends to be rendered in the system [4, 5].

Our interest is in the pressure response of the magnetic-electronic behavior of these ferrous spinels, up to the current state-of-the-art megabar (100 GPa ≈ 0.6 eV $Å^{-3}$) regime. Such energy density input is necessary to impact on the typical pertinent large energy scales (~eV) of important electron interactions in these transition metal oxide systems. Moreover spinels under pressure are known to transit through various structural phases which have been investigated primarily by x-ray diffraction and Raman spectroscopy, for a review see ref [6]. But little is know about the magnetic-electronic aspects of these high pressure structural phases. We have recently addressed this in ferrous spinels in our investigations of $FeCr_2O_4$ [7] and $Fe_2TiO_4$ [8], summarized in the following. We extend similar such investigations to $FeAl_2O_4$ in this study.

In our recent high pressure study of the normal spinel $(Fe)^{IV}(Cr_2)^{VI}O_4$ in which both *A* and *B* cations have a magnetic moment and resultant interesting magnetic-electronic properties (e.g., dynamical Jahn-Teller effect, orbital ordering, multiferroicity), we have demonstrated that an unusual pressure response occurs [7]. A cubic to tetragonal transition occurs at ~12 GPa at room temperature and then persistent increased tetragonality prevails up to nearly a megabar. Fe $\leftrightarrow$ Cr site inversion is onset at ~30 GPa at *room temperature.* This leads to a situation where $Cr^{3+}$ ($3d^3$) is unusually stabilized in tetrahedral coordination. The inverted $Fe^{2+}$, at distorted octahedral sites, then undergo high- to low- spin crossover (HS $\rightarrow$ LS transition) onset at ~60 GPa. Near to 100 GPa 50% of the Fe is LS implying pressure driven inversion to attain $x$ ~ 0.5 in $\left(Fe^{2+}_{1-x}Cr^{3+}_{x}\right)^{IV}\left(Fe^{2+}_{x}Cr^{3+}_{2-x}\right)^{VI}O_4$ . In spite of a more than 30% decrease in unit cell volume the tetragonal high pressure phase remains persistently non-metallic.

In our subsequent study in ulvöspinel $(Fe)^{IV}(FeTi)^{VI}O_4$ [8], a sequence of structural transitions cubic $\rightarrow$ tetragonal $\rightarrow$ orthorhombic (*Cmcm*) $\rightarrow$ orthorhombic (*Pmma*) occur up to ~100 GPa. The symmetry lowering in the orthorhombic post-spinel phase, onset at ~55 GPa, is associated with a disorder/order transition of Fe/Ti at sixfold coordinated sites. Spin crossover is triggered in the *Cmcm* phase and evolves to ~50% abundance of Fe in octahedral sites in the *Pmma* phase at the highest pressure near ~100 GPa , where persistent non-metallic behavior prevails.

The two above-mentioned cases represented some of the first studies of the interplay between magnetic-electronic behavior and structural response of magnetic spinels to high densification. Establishing that pressure driven site-inversion and partial spin crossover may occur, is crucial for knowledge of the physical and chemical properties of such systems in mineral assemblages under extreme pressure-temperature conditions of the deep Earth. Moreover there is also now some intrigue as to the resilience of the charge gap at such high densification. As a follow-up to these recent studies we have conducted similar structural and magnetic-electronic investigations of $FeAl_2O_4$ to the megabar regime. Our study aims to answer questions of whether pressure induces site-inversion, orbital moment quenching, spin state crossover and whether the band gap prevails to such high pressures. This may then constitute a comparative study with the behavior of other ferrous analogs $FeCr_2O_4$ (normal spinel) and $Fe_2TiO_4$ (inverse spinel).

To this end we employ $^{57}$Fe Mössbauer spectroscopy (MS) as arguably the most suitable direct probe of the Fe magnetic-electronic state to extreme conditions, pressures of ~100 GPa and temperatures in the range 300−4 K. For this purpose an $^{57}$Fe isotopically enriched sample was prepared. This was to ensure spectra with an adequate signal-to-noise ratio could be obtained for a reliable analysis, when probing the microscopic pressurized sample cavity to extreme conditions in a diamond anvil cell with MS. As such, in the synthesis procedure briefly described in the next section, some degree of inversion has been attained, $x \sim 0.2$. Hence ~20% of the Fe (tetrahedral) sites on the diamond sublattice are populated by spin vacancies (non-magnetic $Al^{3+}$ cations), and the magnetic ground state is certainly expected to be a spin glass. The pressure response of this spin glass state is thus investigated to very high densification. Complementary x-ray diffraction measurements (XRD) at room temperature and temperature-dependent electrical-resistance measurements, both to the vicinity of a megabar, delineate the structural response to pressurization of the spinel and evolution of the charge gap, respectively.

## II. EXPERIMENTAL

$FeAl_2O_4$ was synthesized using stoichiometric quantities of Fe, $Al_2O_3$ and $Fe_2O_3$. The $Fe_2O_3$ precursor material was enriched to ~20% $^{57}$Fe. The pelleted mixture was heated in an evacuated quartz tube at 1000 °C for 24 hours and then annealed at 500 °C for a week to minimize the inversion. The single-phase purity of the sample, cubic structure and magnetic properties compared with the literature [9] were confirmed by conventional powder XRD and MS at ambient pressure.

The combination of high pressure MS , resistivity and XRD methodologies used here  are very similar to that of our previous investigations of ferrous spinels [7, 8]. Pressure generation was by means of custom built miniature piston cylinder diamond anvil cells (DACs) made at Tel-Aviv University (TAU) [10] for the MS and resistance measurements and 4-pin DACs made at Bayerisches Geoinstitut (BGI) for XRD experiments.  Anvils with culets in the range 200–300 μm diameter have been used for pressurization up to ~100 GPa.  Samples were loaded into 100–150 μm diameter cavities drilled in  rhenium gaskets.  The calibration scales mentioned in ref. [11] were used for pressure determination from the ruby fluorescence measurements. The error in the pressure determination is 5–10% of the reported average pressure from the ruby fluorescence measurements in the case of the MS and resistance experiments.  Liquid argon was used as a pressurizing medium [12, 13] in the MS experiments.

For the resistance measurements the rhenium gasket was insulated with a layer of $Al_2O_3$-NaCl (3:1 *wt. %*) mixed with epoxy to avoid short-circuiting the Pt electrodes used to measure the sample resistance in the DC four-probe configuration.  A few ruby fragments for pressure determination were located in the region between the Pt electrode tips overlapping the sample. No pressure transmitting medium was used, but pressure is effectively transmitted to the sample upon compression by way of the surrounding insulation. Pressure gradients are expected to be small in the distances (20–30 μm) between the tips of the Pt electrodes across which voltage measurements are made.

Variable temperature, 300–5 K *Mössbauer* studies involved $^{57}$Co(Rh) 10-mCi point-source methodology and a top-loading liquid-helium cryostat [14]. Typical data collection time for a single spectrum was ~24 hours. Spectra were analyzed using appropriate fitting programs from which the hyperfine interaction parameters (internal magnetic field $H_{hf}$, isomer (centroid) shift IS and quadrupole splitting/shift QS) and the corresponding relative abundances of the spectral sub-components were derived. The isomer shift in the present work is calibrated relative to α-Fe.

For variable low temperature *resistance measurements* the DAC was placed on a probe connected to a “dip-stick” stepper-motor assembly, which by computer control slowly changed the height of the DAC above the cryogen level in a liquid nitrogen or helium Dewar. The temperature was monitored by a Lakeshore Si (DT-421-HR) diode in proximity to the DAC.

*Powder XRD* measurements were carried out at room temperature up to 80 GPa at the 13ID-D beamline of the Advanced Photon Source (Argonne , Il, USA), with a wavelength of $\lambda$ = 0.3344 Å in angle-dispersive mode and patterns were collected using a MAR CCD detector. Ne was used as the pressure transmitting medium. Pressure was determined from the ruby fluorescence spectra up to 50–60 GPa. The Ne equation of state was used to ascertain pressures through most of the pressure range, and especially at pressures beyond $P$ > 50 GPa [15]. Both methods give rather similar pressure values with the difference not exceeding 0.5 GPa. The error in the pressure determination is about 2% of the reported average pressure. Diffraction images were integrated using the FIT2D [16, 17] and DIOPTAS software [18] . Powder diffraction patterns were analyzed using the GSAS-II software [19] to extract the unit-cell parameters.

The intensities of the diffraction peaks are affected by instrumental and grain-size issues (diamond x-ray absorption and low statistics in the random distribution of sample crystallites). Therefore the Rietveld refinement of the powder diffraction patterns did not result in a good enough fit. Hence diffraction patterns were analyzed by using the whole profile fitting (Pawley) method [20]. Diffraction spectra mainly exhibit peaks from the sample, as well as minor peak contributions from $Al_2O_3$ impurity and the pressure medium (Ne). These phases were considered in the refinements. The reliability factor $R_{wp}$ obtained in the refinement of each of the powder diffraction patterns is ~1%.

## III. RESULTS AND DISCUSSION

Our MS spectra of $FeAl_2O_4$ at ambient pressure involving measurements at room temperature (RT) and liquid helium temperatures are similar to those of Dormann *et al.* [9], see Fig. 1. Dormann *et al.* [9] obtained an inversion factor of $x$ ~ 0.2 from neutron diffraction measurements. They also determined the spin-freezing temperature to be at $T_g$ ~ 13−14 K from the maximum in AC-susceptibility measurements and temperature evolution of the internal magnetic fields in MS experiments. From the similarity of our spectral profiles to that of Dormann *et al.* [9] , we infer that our sample is also a partially inverse spinel, with 20–25% of the $Fe^{2+}$ cations located in *B* (octahedral) sites. We expect the quadrupole doublet splitting (QS), representing deviations of the Fe local environment from cubic symmetry, of *B* sites to be larger than the *A* sites. The latter fourfold coordinated sites normally deviate less from cubic symmetry than the octahedral *B* sites in spinels [21, 22]. Hence the spectrum at ambient conditions has been fitted by using two sub-spectra with an abundance of 80:20 for the different *A*– and *B*- site QS components, respectively, see Fig. 1. The spectrum is not sufficiently well resolved to permit more reliable detailed fitting, e.g., with QS distributions. So broadened sub-spectrum linewidths mimick the QS distribution associated with the various Fe local next-nearest neighbor environments. The line width for the tetrahedral sites is broader than that of the octahedral sites, reflecting the different QS distributions for the two sites.

We have analyzed the magnetic spectra at low temperatures (LT) using a similar fitting strategy to that of Dormann *et al.* [9] for dealing with their temperature dependent MS spectra at ambient pressure where complex spectral profiles of a spin glass occur. A minimum number of multiple sub-spectra involving broadened magnetically split sub-components have been used. This mimicks internal magnetic field ($H_{hf}$) distributions associated with the varied Fe local environments, since we are not able to make a unique identification of the distribution [23]. The $H_{hf}$ values of the discrete sub-components approximate the average values of the true $H_{hf}$ distributions. We have also attempted to maintain consistency between RT and LT spectra for the respective abundances assigned to *A*- and *B*-sites. The hyperfine interaction (HI) parameters of the various sub-components like the centroid IS and $H_{hf}$ values as estimates of average values of $H_{hf}$ distributions are quite reliable. This is considered a good starting model for monitoring changes in the HI parameters and component abundances as evolved at pressure. The pressure evolution of spectra at RT and LT is depicted in Fig.1. Fig. 2 presents the derived HI parameters of the RT spectra in the different structural regimes. The various structural phases are discerned in the XRD measurements depicted in Fig. 3 and derived structural parameters in Fig. 4. The interplay between magnetic-electronic behavior and structural response is discussed in the following.

### A. Low pressure high-spin spinel phase, 1–16 GPa

In this pressure range the XRD data shows that $FeAl_2O_4$ is in the spinel phase, see Fig. 3. The plot of MS HI parameters at RT in Fig. 2 shows that there is a steep increase of QS site-asymmetry parameters up to 10 GPa, indicative of the local distortion of the cubic field. Similar behavior of the QS increase was also observed in $FeCr_2O_4$ where it was attributed to pressure effects on the dynamical Jahn-Teller distortion [7]. The IS values have a negative linear dependence on $\rho_s(0)$, the *s*-electron density at the nucleus [24, 25] . Thus the decrease in IS values upon increasing pressure is anticipated as the *s*-electron density increases under compression.

Low-temperature MS spectra, measured at ~10 K, each retain the same profile as ambient pressure up to 10 GPa. From this we infer that the spin glass nature, reported by Dorman *et al.* [9], remains as such in this pressure range. It should be noted that it is the combination of complex magnetic Mössbauer spectra and the peak features in the AC-suceptibility measurements of Dormann *et al.* [9] and Soubeyroux *et al.* [26] , respectively, that established the occurrence of spin-glass behavior in an $FeAl_2O_4$ sample with inversion factor $x$ ~ 0.2 at ambient pressure. Up to ~10 GPa the LT spectral profiles of our sample, similar to that of Dorman *et al.* [9], are typical of magnetic spectra which develop complexity manifested in a distribution of $H_{hf}$ values below a temperature $T_g$ characteristic of spin glass freezing. This would primarily be directly deduced from temperature-dependent AC-susceptibilty measurements. Hence similar to Dormann *et al.*, the spectra at LT and ambient pressure in Fig. 1 are fitted with one component having a conspicuously large magnetic field of ~48 T (~20% abundance, attributed to inverted Fe at *B*-sites). Additional multiple (three) magnetically split components with $H_{hf} \leq 25$ T are used, as a representation of the $H_{hf}$-distribution associated with the varied local environments and magnetic interactions of Fe at *A*-sites.

The MS spectra show magnetic hyperfine splittings as temperature rises to above 25 K upon increasing pressure to beyond ~10 GPa. It is unlikely that a magnetic hyperfine splitting at such high temperatures is a signature of spin-freezing at such elevated temperatures in a spin glass. In which case we suppose that at these reduced interatomic spacings a different magnetic ground state has been stabilized compared to ambient pressure.

The assignment of the large $H_{hf}$ ~ 48 T component to Fe at *B*-sites (inversion) is partly based on how it evolves at much higher pressures, as discussed in the following sub-sections.

### B. Tetragonal spinel phase, 18–55 GPa

The XRD profiles in Fig. 3 indicate that beyond 16 GPa tetragonality occurs and prevails as such to ~55 GPa. Fig. 4 depicts lattice parameters and unit-cell volume as a function of pressure. The behavior of the unit-cell volume shows an "apparent stiffening" up to ~35 GPa, after which there is a change to a much steeper pressure dependence of the unit-cell volume (increased compressibility) up to ~55 GPa where this phase has disappeared, see Fig. 4(b). For comparison purposes we extrapolate the equation of state (EOS) of the low pressure cubic phase to well beyond 16 GPa up to ~50 GPa. This may be compared with the structural response of cubic and tetragonal phases in other normal spinels, e.g., $ZnGa_2O_4$ [27]. Therein there is no discernable change in compressibility associated with the cubic $\rightarrow$ tetragonal transition. Therefore the compression of tetragonal $FeAl_2O_4$ may be deemed as highly irregular and merits deeper consideration, as discussed in the following.

A representative Mössbauer spectrum of the tetragonal phase at 30 GPa and RT is shown in Fig. 1. The abundance of the large QS component attributable to sixfold (octahedral) coordination, has increased appreciably ~40%, compared to that in the cubic phase. This prevails to ~50 GPa where further conspicuous changes then become onset. The irregular compression behavior of the unit-cell volume in conjunction with the increase in abundance of the large QS component of the spectrum at RT is considered as evidence of pressure-induced Fe $\leftrightarrow$ Al *site inversion triggered at room temperature*. Thus in Fig. 4(b) the up-turn in unit-cell volume to ~40 GPa signifies the progression of inversion to completion, after which normal compressibility behavior occurs.

The LT magnetic spectrum at 30 GPa in Fig. 1 also shows some changes to the profile in the centralized velocity region near zero mm $s^{-1}$. The spectrum is characterized by a large $H_{hf}$ ~ 46 T component and 20−25 T discrete components equivalent of $H_{hf}$-distribution averages. The originally inverted Fe sites are represented by the large $H_{hf}$ ~ 46 T component. Further inversion and its magnetic signatures are convoluted into the rest of the $H_{hf}$-distribution by an equivalent discrete ~25 T component as manifested in some changes in the centralized region compared to the cubic phase. The third component, a small ~20 T $H_{hf}$-distribution average equivalent, is ascribed to the varied local environments of fourfold coordinated sites HS(IV).

It is also noteworthy that the pressure dependence of $R$(300 K) shows non-monotonic behavior in the 20−50 GPa range, see Fig. 5. The value of $R$(300 K) changes by only one order of magnitude from its initial value up to 50 GPa. This also translates to hardly any change in the activation energy of the original insulating state. The top right hand inset of Fig. 5 also shows plots of the data in terms of Arrhenius thermally activated hopping behavior, $\ln R \propto 1/T$ . Below ~41 GPa linearity is only in a limited high temperature range from which the activation energy $E_a$ is extracted. At higher pressures $P \geq 52$ GPa linearity occurs throughout the temperature range 300−4 K. This indicates that a temperature induced change in conduction mechanism which occurs at lower pressures, does not occur at higher pressures. This could be associated with the Fe $\leftrightarrow$ Al site inversion occurring at $P < 40$ GPa.

In the vicinity just prior to ~50 GPa XRD patterns show evidence of reflections of another structural transition occurring, see Fig. 3 (a). Based on the evolution of this new phase to higher pressures $P > 55$ GPa, we have deduced that this is the orthorhombic *Cmcm* post-spinel phase. Thus XRD data indicates that 46−55 GPa is a regime of coexistence of tetragonal spinel and orthorhombic post-spinel phases. $FeAl_2O_4$ is fully converted to post-spinel as pressure rises above 55 GPa.

At ~55 GPa RT and LT MS profiles show rather conspicuous changes compared to spectra at lower pressure. The RT spectrum develops pronounced asymmetry in the overall spectral envelope and the LT spectrum exhibits an appreciable increase in intensity in the centralized velocity region. These are considered manifestations of a new doublet component emerging. In the RT spectrum this is a component with HI parameters of IS = 0.45 mm $s^{-1}$ and QS = 0.8 mm $s^{-1}$ . This new doublet component remains as such down to the lowest temperature measured, 12 K. This indicates that it is non-magnetic in nature and accounts for the increased intensity in the centralized region of the magnetic spectrum, see Fig. 1. This diamagnetic component coexists with other sub-spectra which may be remnants of the low pressure phase as well as with anticipated eightfold coordination sites (HS(VIII)) of the post-spinel phase.

These HS(VIII) sites likely evolve from the tetrahedral HS(IV) sites of the tetragonal spinel, involving the reconstructive mechanisms described by Arévalo-López *et al.* [28].

### C. *P* > 55 GPa, partially spin converted post-spinel

The XRD data in Fig. 3 indicates that above 55 GPa only the post-spinel phase is discerned. There is a much steeper change in unit-cell volume $\Delta V/V$ ~ 10% preceding this, in the range 45−50 GPa. In addition to the normal compressibility, we attribute this to both the transition to a denser orthorhombic post-spinel structure and also to what the emergence of a diamagnetic spectral component in the MS spectra is suggesting. This non-magnetic component has HI parameters typical of low-spin (LS) Fe at sixfold coordinated sites, from 3*d* spin crossover at HS(VI) sites [24]. Spin crossover and the consequential 3*d* electron spin-pairing HS:(↑↑↑↓)(↑↑) ⇒ LS:(↑↓↑↓↑↓)( ), $t_{2g}^4 e_g^2 \rightarrow t_{2g}^6 e_g^0$, leads to a change of atomic spin $S$=2 moment to a collapsed $S$=0 diamagnetic ($H_{hf}$=0) state. The LS electronic configuration also represents a more symmetric local environment of surrounding charge. Consequently the QS (site-asymmetry) parameter values are normally much lower than in the HS state, as discerned by the LS(VI)-designated values in the *Cmcm* section of Fig. 2. The other tell-tale signature of an HS → LS change is the pronounced decrease in IS, Fig. 2 bottom panel. This is because of the change in 3*d*-electron shielding of *s*-electrons, from the re-distribution of charge in the 3*d* orbitals. This in turn affects IS because of its relation to the *s*-electron density, as delineated at the end of the first paragraph in sub-section A [24].

We identify these as the Fe sites evolved from Fe ↔ Al site inversion. Recall that such sites originated in the original sample synthesis process (20−25% abundance) and additional inversion is triggered at high pressure in the tetragonal phase (20−25% abundance). There is also an anticipated volume contraction associated with the HS → LS spin crossover process [29, 30], due to the redistribution of charge from $e_g$ to $t_{2g}$ orbitals in ferrous iron $t_{2g}^4 e_g^2 \rightarrow t_{2g}^6 e_g^0$. The QS site-asymmetry parameters in Fig. 2 for the various local environments show an increasing trend. This is presumably indicative of appreciable polyhedral distortions at this high compression, resulting in increased crystal field (CF) values that trigger spin crossover at octahedral sites [30].

Differences in the onset pressures of spin crossover as discerned by XRD (45−50 GPa) and MS (50−55 GPa) may be attributed to the different pressure transmitting media used and how the degree of non-hydrostaticity affects the electronic transition [31]. In addition, in the synchrotron XRD measurements the signal derives from a small central part of the sample. Whereas in Mössbauer pressure studies the signal is collected from a much larger ~2⁄3 inner region of the sample diameter, resulting in possible pressure gradient effects which could be significant in determining phase transition pressures. Differences may also arise from the higher sensitivity of the synchrotron XRD method, where the higher signal/noise ratio allows the detection of new components at a level where its relative abundance is too small for discernment by MS in a spectrum with strongly overlapping sub-components.

Thus the pronounced volume change $\Delta V/V$ ~ 10% in the vicinity of ~50 GPa evident in the unit-cell volume evolution in this pressure regime, Fig. 4(b), is partially attributable to both the tetragonal → post-spinel structural change and emergence of LS states, LS(VI), at sixfold coordinated sites in the post-spinel phase as originated from site-inversion processes (involving HS(VI)). Normally such a densification should manifest as corresponding decreases in IS values at RT of the MS sub-components in Fig. 2 [25]. At HS(VIII) sites this is compensated by the increase in coordination (IV) → (VIII) which normally leads to an increase in IS [32]. The large difference in IS for HS(VI) and LS(VI) likely has contributions from both the concurrent spinel/post-spinel densification as well as from the effects of spin-pairing [29, 33].

The evolution of the MS spectra beyond ~55 GPa , Fig. 1, indicates that the abundance of the LS component progressively increases with a corresponding diminution of HS(VI) abundance. Beyond a megabar , at ~106 GPa, the LS abundance is ~50% as deduced from the (shaded) absorption areas of the sub-components in Fig. 1.

A broadened magnetic component with average $H_{hf}$ ~33 T is still present in the LT spectrum at the highest pressure attained. This is ascribed to the 50% abundant Fe sites in eightfold coordination, HS (VIII), originated from non-inverted HS(IV) sites of the cubic spinel phase. In such a high-coordination local environment the crystal field splitting is not large enough for spin crossover to ensue [8, 34].

Comparison of the $H_{hf}$-distribution average equivalent for the HS(VIII)-designated component indicates that it has increased from ~20 T at 55 GPa to ~33 T at 106 GPa. These are depicted as the dotted line sub-components in Fig. 2. In ferrous iron $H_{hf} = H_C + H_L$, comprised of an *s*-electron spin-density related Fermi-contact term $H_C$ and orbital moment term $H_L$ which may have opposite signs [35]. Modifications to the 3*d* level scheme due to compression changes in the eightfold coordinated local environment, lead to quenching of $H_L$ and consequently to an increase in $H_{hf}$ [8]. In the LT spectrum at 68 GPa, two dotted line sub-components with different $H_{hf}$ average values are required to make up sufficient absorption area so as to be compatible with the HS(VIII) component area at RT. At this intermediate pressure some HS(VIII) sites have a higher degree of $H_L$ quenching than others, likely due to inhomogeneous distortions at eightfold coordinated sites. At the highest pressure all HS(VIII) sites have $H_L$ quenched to the same extent such that an average $H_{hf}$ ~ 33 T component is discerned.

The final picture that emerges of the high density $FeAl_2O_4$ post-spinel phase beyond a megabar, is that of a partially spin-converted lattice structure. About half of the $Fe^{2+}$ reside in sixfold coordinated sites from prior inversion processes and all of this has undergone spin crossover. The remainder reside in eightfold coordinated sites and still exhibit evidence of a magnetic ground state. The post-spinel sublattice involving eightfold coordinated sites has both $Fe^{2+}$ (HS(VIII)) and $Al^{3+}$ constituents. Similarly the sublattice involving sixfold coordinated sites has LS(VI) $Fe^{2+}$ and $Al^{3+}$ constituents. This renders the overall lattice configuration as one involving a large degree of site disorder.

The resistance measurements up to the highest pressure of ~90 GPa in Fig. 5 indicate that $R$(300 K) decreases by less than two orders of magnitude over the entire pressure range. At the highest pressure there is also a tendency toward plateau behavior, typical of what has been seen in other (partially) spin converted lattices involving ferrous iron analogs [7, 8]. A high density ~400 meV gapped state prevails to ~100 GPa as inferred from an extrapolation of the activation energies from lower pressures.

The ~26% decrease in unit-cell volume up to a ~100 GPa from combined lattice compressibility, structural transitions and partial spin crossover should lead to an appreciable increase in 3*d* band-broadening and gap closure, yet the compound remains resiliently non-metallic. We rationalize this in terms of an increase in $U_{eff}$ ,

the electron-correlation gap between upper and lower Hubbard sub-bands, at spin crossover which is specific to $Fe^{2+}(3d^6)$ [36] , as we have suggested for the other ferrous analogs $FeCr_2O_4$ and $Fe_2TiO_4$ [7, 8]. Additionally we also suggest Anderson localization effects may be prevalent from disorder as a result of both Fe/Al site inversion and the mixed high-spin/low-spin electronic configurations at Fe sites in the $\left(Fe^{2+}_{1-x}Al^{3+}_{x}\right)^{VIII}_{HS}\left(Fe^{2+}_{x}Al^{3+}_{2-x}\right)^{VI}_{LS}O_4$ , x ~ 0.5, lattice structure.

It is noteworthy that according to XRD data in Fig. 4(b) the transition from the tetragonal to orthorhombic (*Cmcm*) structure is irreversible. Upon decompression down to at least 28 GPa no phase transition back to the intermediate $I4_1/amd$ structure was observed. Furthermore, we do not observe a steep change in unit-cell volume at 40–50 GPa, similar to that observed upon compression, corresponding with the HS-LS transition of $Fe^{2+}$ in octahedral sites (Fig. 4(b)). There is only a very sluggish deviation of the $V(P)$ curve from the high pressure *Cmcm* EOS extrapolated down to 28 GPa (Fig. 4(b), dashed blue line), which reaches ~2% at 28 GPa. Such a volume deviation can be related to a spin-state change back to HS at one quarter of the Fe sites in the structure. This takes into account an estimate of the relative difference in octahedral volumes of ~21% expected for HS and LS $FeO_6$ polyhedrons [8] and that Fe, constituting one third of the cations, is equally distributed at sixfold coordinated (VI) sites where a spin-state change is possible and eightfold coordinated (VIII) sites.

## IV. SUMMARY AND CONCLUSIONS

Hercynite $FeAl_2O_4$ up to a megabar has the following sequence of structural transitions : cubic ( $Fd\overline{3}m$ ) $\rightarrow^{16\ \mathrm{GPa}}$ tetragonal ($I4_1/amd$) $\rightarrow^{45\ \mathrm{GPa}}$ orthorhombic (*Cmcm*), at the indicated onset transition pressures. The following physical properties of these various structural phases are manifest :

(i) In the cubic spinel spin ordering temperatures rise from a spin-freezing $T_g \sim 14$ K at ambient pressure to ordering temperatures that exceed 25 K above 10 GPa. In the tetragonal phase the compression of the unit-cell volume is highly irregular. This is ascribed to “cold compression” driven Fe ↔ Al site inversion at room temperature.

(ii) There is a steep change in unit-cell volume $\Delta V/V \sim 10\%$ in a pressure window of 10 GPa encompassing ~50 GPa, to culminate in the fully stabilized *Cmcm* post-spinel phase as pressure rises above 55 GPa. Mössbauer spectra also evidence the emergence of a diamagnetic spectral component. Part of this volume contraction as well as the diamagnetic spectral component are accounted for by spin crossover at sixfold coordinated sites, initiated at the end of the pressure range of the tetragonal phase and progressed in the emergent post-spinel phase.

(iii) These are Fe sites originated from prior inversion processes from both sample synthesis conditions and pressure effects on the tetragonal phase. The low-spin abundance evolves to maximally ~50% of the Fe sites in the post-spinel phase at the highest measured pressure of 106 GPa. This high density post-spinel is inferred to have the formulation $\left(Fe^{2+}_{1-x}Al^{3+}_{x}\right)^{VIII}\left(Fe^{2+}_{x}Al^{3+}_{2-x}\right)^{VI}O_4$ with $x$ ~ 0.5 and is a highly disordered lattice structure.

(iv) The structural transition from tetragonal (*I*4$_1$/*amd*) to orthorhombic (*Cmcm*) is irreversible, possibly as a result of the appreciable disorder in the *Cmcm* phase. Furthermore, upon decompression to 28 GPa there is an increase in unit-cell volume with no change in symmetry. This deviates to above the EOS of the orthorhombic *Cmcm* high pressure phase from a back transition to HS, limited to about half of all the Fe in octahedral sites (partial hysteretic behavior of the electronic HS-LS transition).

In closing : partial spin crossover appears to be the preferred electronic response of ferrous spinels at sufficiently high density, as demonstrated in our work on $FeCr_2O_4$ (chromite), $Fe_2TiO_4$ (ulvöspinel) and again in the current case of $FeAl_2O_4$ (hercynite). In the chromite and hercynite normal spinels at room temperature pressure drives inversion, notably HS(IV) $\rightarrow$ HS(VI) site diffusion. Appreciable polyhedral distortions and increased CF values trigger spin crossover at selected (VI) sites in high-density phases of these partially inverse systems. Similar site-specific spin crossover occurs in the originally fully inverse ulvöspinel system to render a partially spin converted post-spinel phase. It is noteworthy that in all of the ferrous iron systems we have investigated, this partial HS-LS transition takes place in the vicinity of ~50 GPa, irrespective of whether the structure is tetragonal or orthorhombic. This is a regime involving sufficiently reduced and distorted octahedral volumes such that the corresponding CF is at critical values to invoke spin crossover [30].

At a megabar these ferrous iron systems are all spin-diluted by "Hund's spin vacancies" (LS sites), to differing degrees of dilution. In spite of this and the appreciable 3*d* band broadening from various lattice responses that contribute to the unit-cell reduction, they continue to have concurrent resilient magnetic and non-metallic ground-state features of strong electron correlations. It is likely that the positive pressure dependence developed by $U_{eff}$ at spin crossover, specific to the $3d^6$ electronic structure [36], prevents the anticipated break down of strong electron correlations anticipated at such high densities.

## ACKNOWLEDGEMENTS

This research was supported by the Israel Science Foundation (Grant No. 1189/14). G.R.H. acknowledges financial support from the National Research Foundation of South Africa (Grant No. 105870). We acknowledge M. Shulman for his help in DAC preparation and XRD experiments. Portions of this work were performed at GeoSoilEnviroCARS (The University of Chicago, Sector 13), Advanced Photon Source (APS), Argonne National Laboratory. GeoSoilEnviroCARS is supported by the National Science Foundation - Earth Sciences (EAR-1634415) and Department of Energy-GeoSciences (DE-FG02-94ER14466). Use of the COMPRES-GSECARS gas loading system was supported by COMPRES under NSF Cooperative Agreement EAR-1606856 and by GSECARS through NSF grant EAR-1634415 and DOE grant DE-FG02-94ER14466. This research used resources of the Advanced Photon Source, a U.S. Department of Energy (DOE) Office of Science User Facility operated for the DOE Office of Science by Argonne National Laboratory under Contract No. DE-AC02-06CH11357.

## FIGURE CAPTIONS

FIG. 1. Selected Fe Mössbauer spectra at room temperature (RT, left panel ) and low temperatures (LT, right panel), in the pressure regimes discussed in the text. Solid line through the data points represent the overall fit to the data from the sum of sub-components shown. Dotted line sub-components refer to the fourfold (IV) or eightfold (VIII) coordinated sites and other sub-components refer to sixfold (VI) coordinated sites, discussed in the text. Multiple magnetically split components representing the $H_{hf}$ distribution associated with *A*-sites (IV) of the complex spectrum of the spin glass state at LT and ambient pressure [9], are not shown. The shaded component represents low-spin Fe sites originated from (IV) $\rightarrow$ (VI) site-inversion processes.

FIG. 2. Pressure dependences of the hyperfine interaction parameters from best fits to the Mössbauer spectra at RT. QS and IS parameters from fitted sub-components of RT spectra exemplified in Fig.1 are in the top and bottom panels, respectively. Three pressure regimes of structurally distinct phases demarcated by the vertical dotted lines have been identified and discussed in the text. Roman numerals IV, VI and VIII refer to four- six- and eightfold coordinated tetrahedral, octahedral and bicapped trigonal prismatic sites, respectively. Spectra in the range ambient–18 GPa, where XRD discerns cubic behavior at 300 K, are best fit at this temperature with two QS components having similar IS values as discussed in the text. These are associated with high-spin Fe local environment distributions of sixfold coordinated Fe sites from inversion HS(VI) and fourfold coordinated sites HS(IV). Spectra in the regime 20–55 GPa are deconvoluted in a similar way. Beyond 50–55 GPa the structural transformation of fourfold to eightfold coordinated HS(VIII) sites and evolution of the sixfold coordinated sites across the $I4_1/amd \rightarrow Cmcm$ reconstructive transition are described in ref. [28]. Parameters for the low-spin sites are delineated by the (LS) labels.

FIG. 3. (a) Pressure evolution of the XRD patterns up to ~80 GPa ($\lambda$ = 0.3344 Å). The three pressure regimes of structurally distinct phases are demarcated on the vertical axis, as discussed in the text. (b) Examples of Pawley method refinements [20] of the XRD patterns at pressure in the various structural phases. Contamination

reflections from minor phase $Al_2O_3$ remnant and solid Ne pressure transmitting medium are indicated. The unit-cell parameters of the phases depicted in the three panels are: cubic $a$ = 8.087(1) Å ; tetragonal $a$ = 5.622(9) Å and $c$ = 7.951(13) Å ; orthorhombic $a$ = 8.703(9) Å, $b$ = 8.688(9) Å and $c$ = 2.654(8) Å.

FIG. 4. Pressure dependences of the lattice parameters (a) and unit-cell volume (b) of the various structural phases. The vertical error bars do not exceed the size of the symbols. Solid lines through the data symbols are fits with a Birch-Murnaghan equation of state (EOS) to obtain bulk modulus $K_0$ and constraining its derivative at $K_0'$=4 [37]. The vertical dashed lines in (b) demarcate the pressures where transitions to new structural phases are completed. Note the irregular behavior of both the lattice parameters and unit-cell volume of the tetragonal phase in the range 20–50 GPa. In particular, note the change in pressure dependences of $c(P)$ and $a(P)$ beyond ~22 GPa followed by a steeper $c$-parameter and $c/a$ ratio decrease above 40 GPa. A steeper decrease of the $c$-parameter is also observed for the emergent *Cmcm* phase up to ~53 GPa. Above 53 GPa the compressibility of the orthorhombic structure is isotropic. This behavior is attributable to Fe ↔ Al site inversion followed by a sluggish HS-LS transition initiating at octahedral sites (VI) as discussed in the text. The EOS of the cubic phase is extrapolated to ~50 GPa to emphasize the anticipated behavior if there is no pressure instigated site inversion [27]. For the decompression data the EOS of the orthorhombic phase is extrapolated to ~28 GPa to reveal unit-cell volume changes related to only partial recovery of the Fe HS state at sixfold coordinated sites.

FIG. 5. Pressure dependence of the resistance at 300 K in the main panel. Top right inset shows linearized temperature dependent data at various pressures assuming Arrhenius activated hopping transport, $\ln(R) \propto \frac{E_a}{k_B T} + const.$ . The activation energy $E_a$ is obtained from the slope of linear fits to these plots to obtain the charge gap for intrinsic conduction $E_g = 2E_a$ . Bottom left-hand inset shows the pressure dependence of $E_a$ , the solid line through the data points is to guide the eye. Dashed vertical lines in the main panel delineate pressures where transitions to new structural phases are completed according to XRD experiments, see text and preceding Fig. 4.

electronic orbital states becomes energetically favorable and HS → LS crossover occurs.

Fig. 1 Xu *et al.*

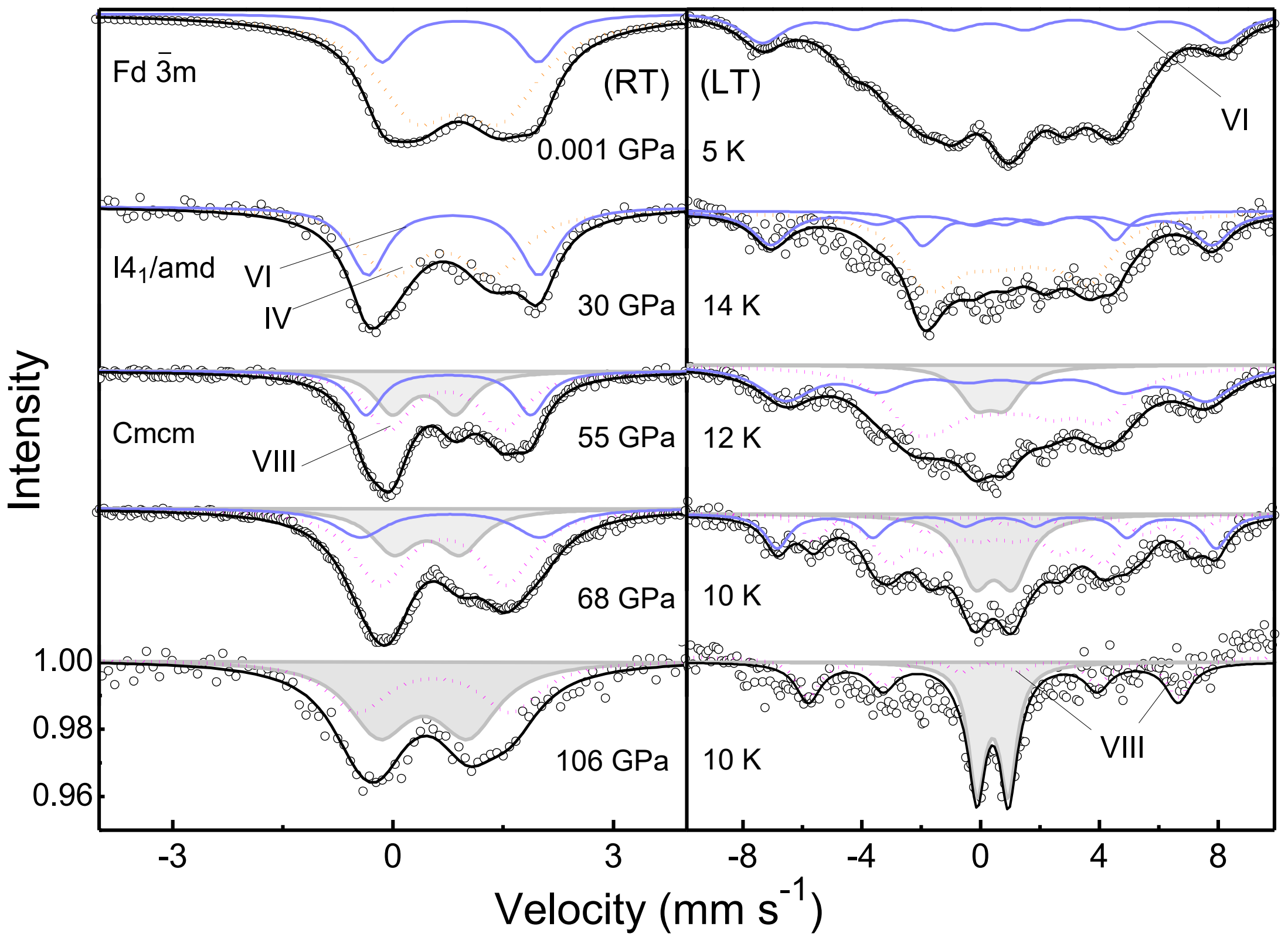

Fig. 2   Xu *et al.*

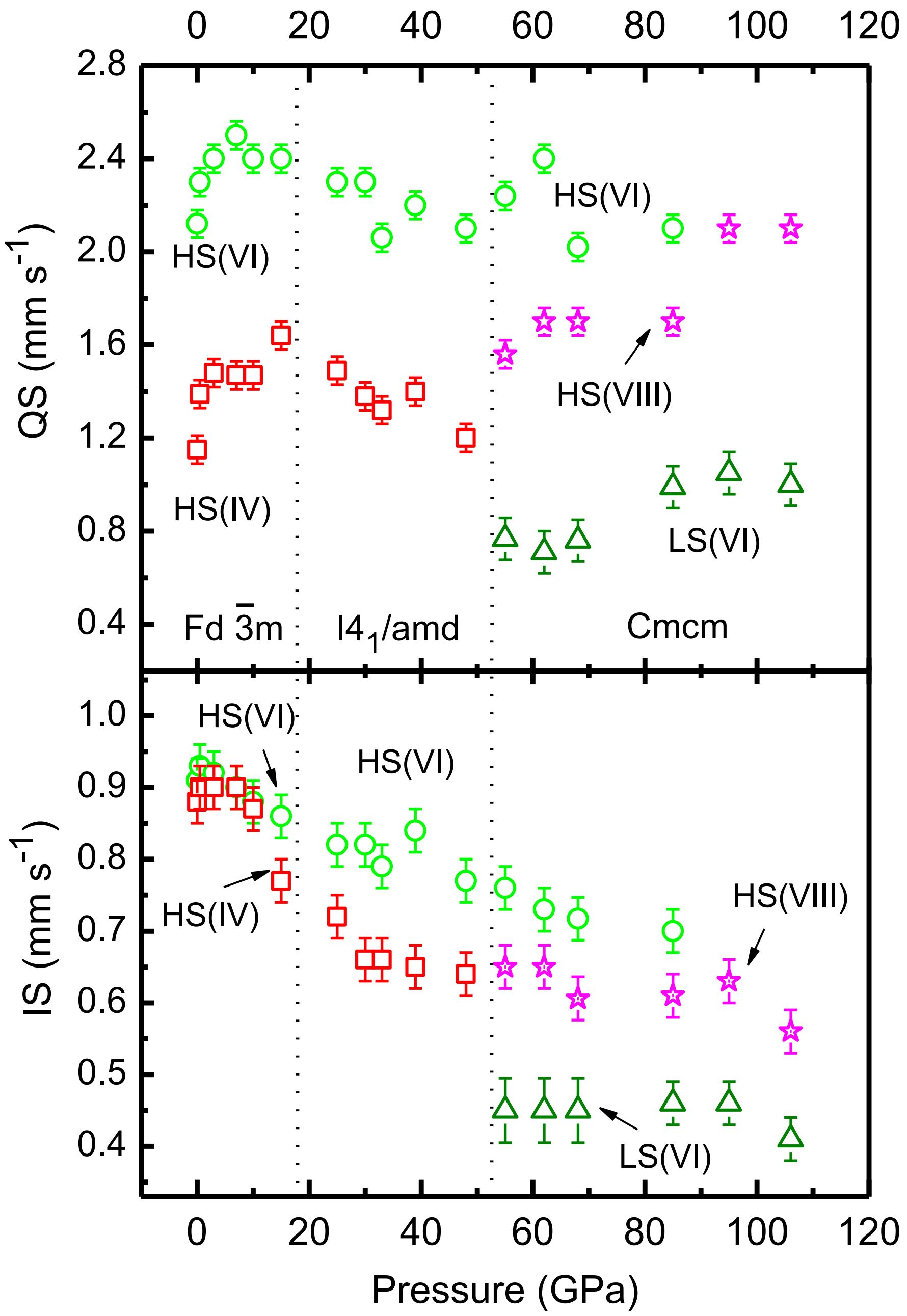

Fig. 3 Xu *et al.*

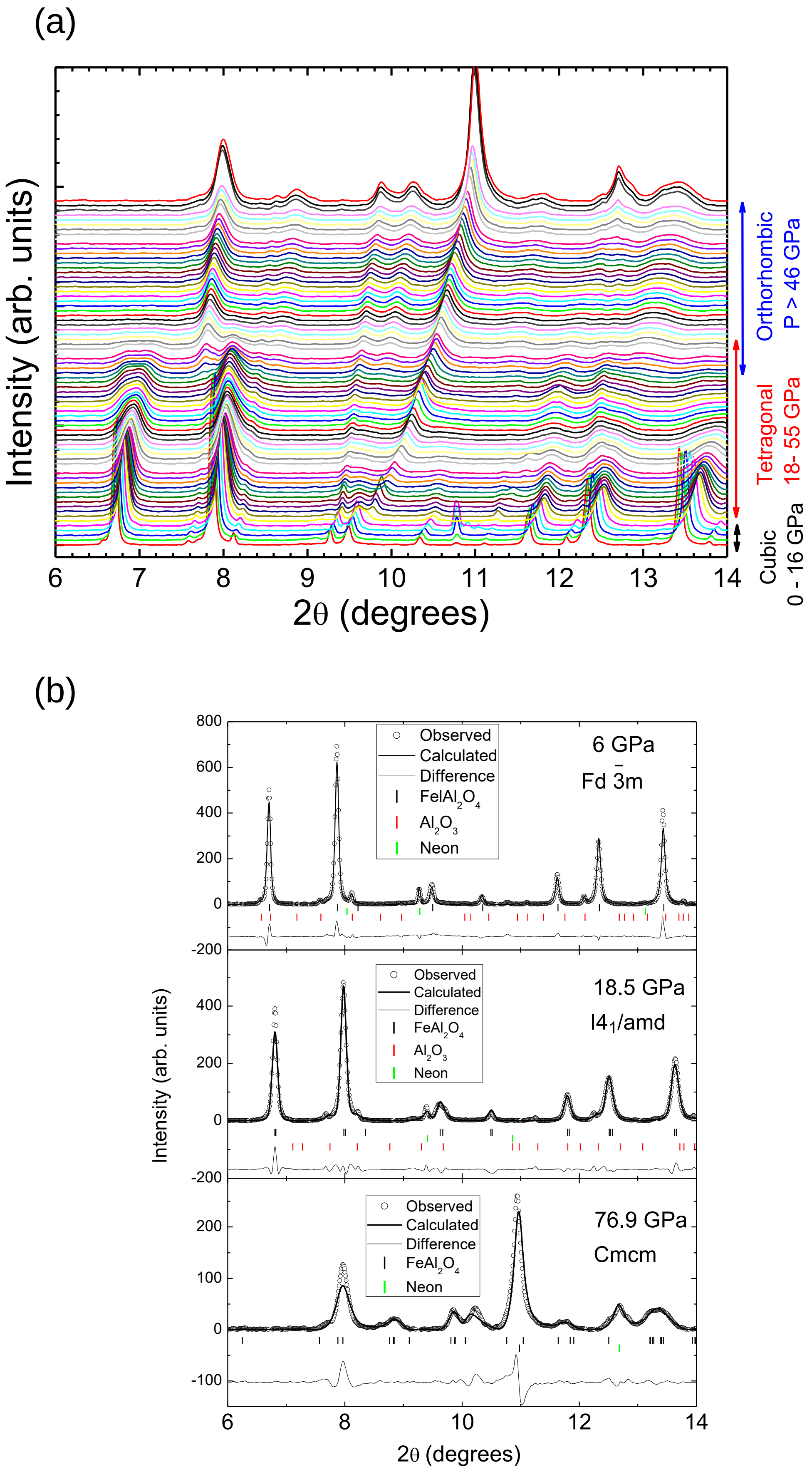

Fig. 4 Xu *et al.*

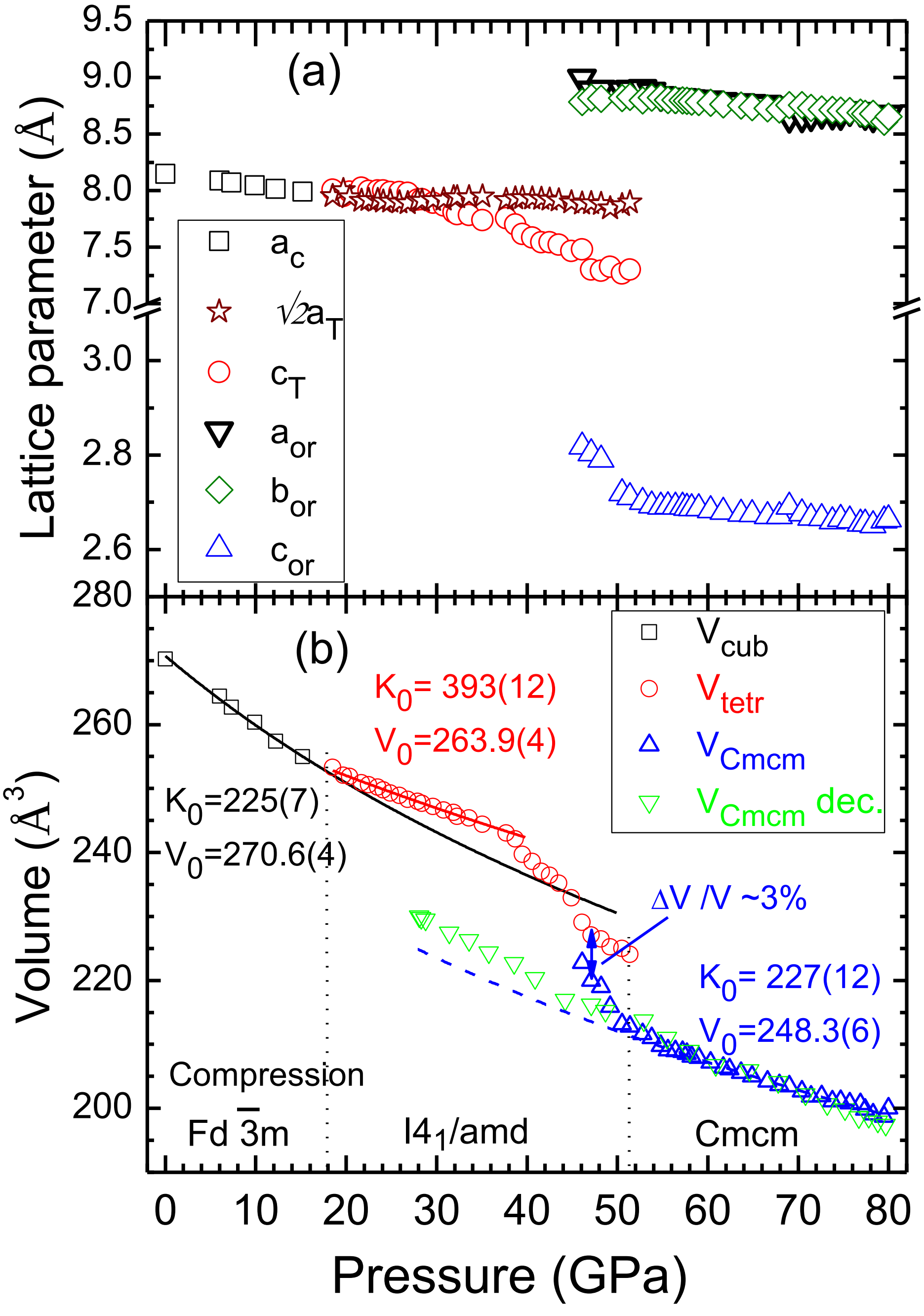

Fig. 5 Xu *et al.*

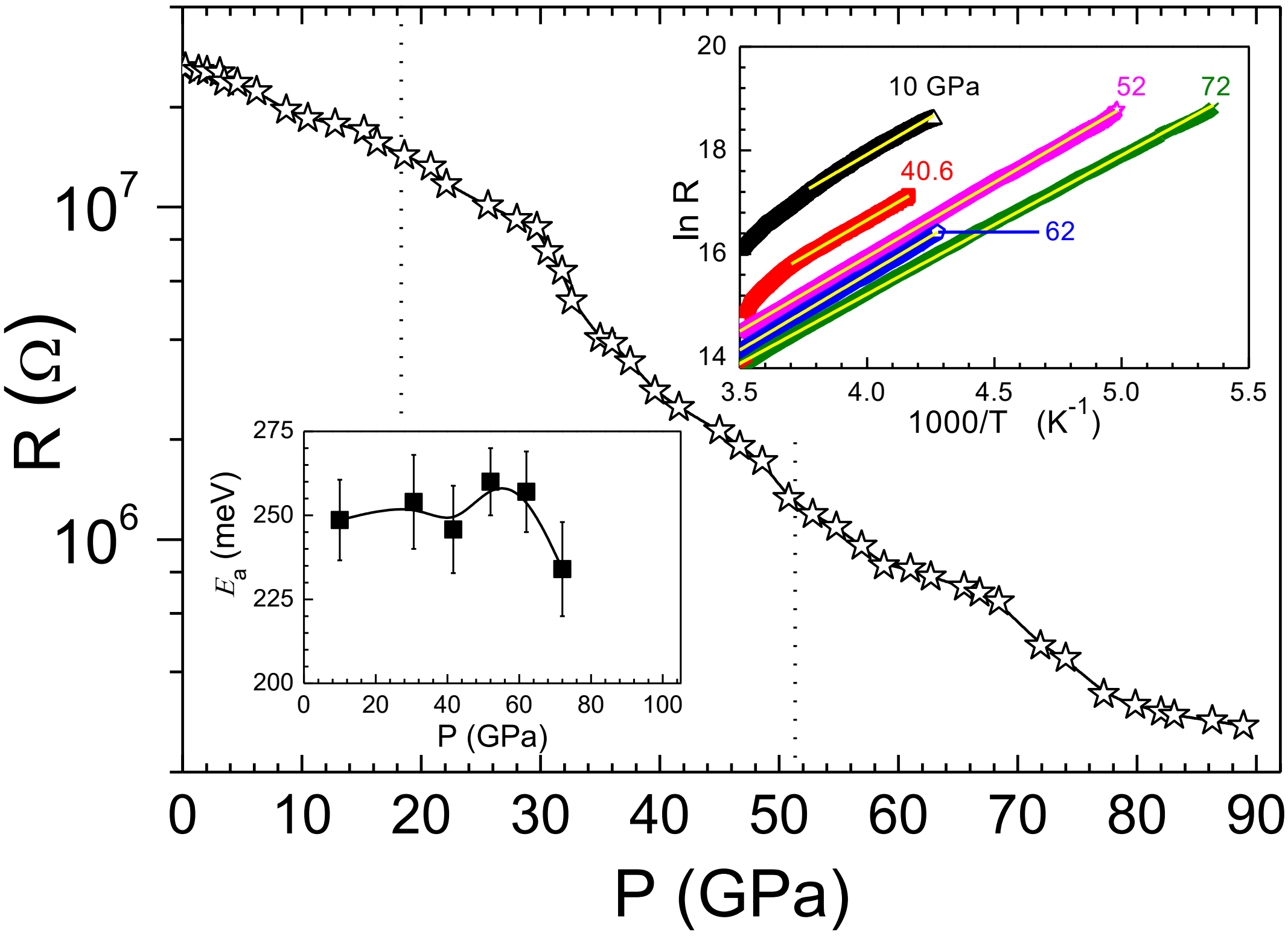